\documentstyle{l-aa-1259}

\input psfig

\renewcommand{\deg}{$^{\circ}$}
\renewcommand{\sec}{"}

\def\jref#1 #2 #3 #4 {{\par\noindent \hangindent=2em \hangafter=1
      \advance \rightskip by 0em #1, {\it#2}, {\bf#3}, #4.\par}}
\def\rref#1{{\par\noindent \hangindent=2em \hangafter=1
      \advance \rightskip by 0em #1.\par}}
\newcommand{\ggg}{$\gamma$}
\newcommand{\eee}{$e^{\pm}$}

\newcommand{\ergs}{\rm \su  erg \su s^{-1}}

\newcommand{\su}{\hspace*{.1in}}
 
\newcommand{\etal}{{et~al.$\,$}}

\def\loe{\lower 0.6ex\hbox{${}\stackrel{<}{\sim}{}$}}
\def\goe{\lower 0.6ex\hbox{${}\stackrel{>}{\sim}{}$}}

\newcommand{\be}{\begin{equation}}
\newcommand{\en}{\end{equation}}

\newcommand{\bo}{ }

\def\aa #1 #2 {A\&A, {#1}, #2}
\def\aas #1 #2 {A\&AS, {#1}, #2}
\def\araa #1 #2 {ARA\&A, {#1}, #2}
\def\mon #1 #2 {MNRAS, {#1}, #2}
\def\apj #1 #2 {ApJ, {#1}, #2}
\def\apjs #1 #2 {ApJS, {#1}, #2}
\def\apjl #1 #2 {ApJ, {#1}, #2}
\def\aj #1 #2 {AJ, {#1}, #2}
\def\nat #1 #2 {Nature, {#1}, #2}
\def\pasj #1 #2 {PASJ, {#1}, #2}
\def\pasp #1 #2 {PASP, {#1}, #2}
\def\nn{\noindent}

\newcommand{\psr}{PSR~B1259-63 }
\newcommand{\psrp}{PSR~B1259-63}

\title{ High-Energy Emission from the PSR~B1259-63  System near Periastron}
\author{M. Tavani\inst{1}
\and J.E. Grove\inst{2}
\and W. Purcell\inst{3}
\and W. Hermsen\inst{4}
\and L. Kuiper\inst{4}
\and P. Kaaret\inst{1}
\and E. Ford\inst{1}
\and R.B. Wilson\inst{5}
\and M. Finger\inst{5}
\and B.A. Harmon\inst{5}
\and S.N. Zhang\inst{6}
\and J. Mattox\inst{7}
\and D. Thompson\inst{8}
\and J. Arons\inst{9} }
\begin{document}
\offprints{M. Tavani}
 
\institute{
{Columbia Astrophysics Laboratory, Columbia University, New
York, NY 10027, USA}
% \\ e-mail: tavani@meleas.phys.columbia.edu }
\and
{Code 7651, Naval Research Laboratory, Washington, DC  20375, USA}
\and
{Department of Physics \& Astronomy, Northwestern University, Evanston, IL 60208
, USA}
\and
{SRON-Utrecht, Sorbonnelaan 2, 3584 CA Utrecht, The Netherlands}
\and
{NASA/MSFC, Huntsville, AL, 35812, USA}
\and
{Universities Space Research Association, NASA/MSFC, Huntsville, AL, 35812, USA}
\and
{Department of Astronomy, University of Maryland, College Park,  MD 20742, USA}
\and
{Code 662, NASA/Goddard Space Flight Center, Greenbelt, MD 20771,USA}
\and
{Department of Astronomy, University of California at Berkeley, Berkeley, CA 94720, USA} }

\date{Received ...........   ; Accepted .......  }
 
\maketitle

% \vskip .3in
 
% \cen{\bf Abstract}
% \vskip .2in

\begin{abstract}

We report the results of a CGRO  multi-instrument 3-week
observation of the 
binary system containing the 47 ms pulsar PSR~B1259-63 orbiting around
a Be star companion in a very eccentric orbit. The PSR~B1259-63 binary
is a unique system  for the study of the interaction of a rapidly
rotating pulsar with time-variable nebular surroundings. CGRO observed
the PSR~B1259-63 system in coincidence with its most recent periastron
passage (January 3-23, 1994). {\it Unpulsed and non-thermal} hard X-ray
emission was detected up to 200 keV, with a photon index $1.8 \pm 0.2$
and a flux of $\sim 4$~mCrab, corresponding to a luminosity of
a few~x~$10^{34}$ erg/s at the distance of 2 kpc.  The hard X-ray flux
and spectrum detected by CGRO agrees with the X-ray emission measured by 
simultaneous ASCA observations.  EGRET  upper limits
 are significant, and exclude strong inverse Compton
cooling in the \psr  system. We interpret the observed  non-thermal emission as
synchrotron radiation of shocked electron/positron pairs of the
relativistic pulsar wind  interacting with the mass outflow from the
Be star.
 Our results %mt4 show
clearly indicate, 
for the first time in a binary pulsar, 
that  high energy emission can be shock-powered
rather than caused by accretion.
Furthermore, the \psr system shows 
that shock acceleration can increase the
original energy  of pulsar wind particles by a factor $\goe 10$, despite
the high synchrotron and inverse Compton cooling rates near periastron. 
We also present results of an extensive search for pulsed
gamma-ray emission from PSR B1259-63. The lack of pulsations constrains
models of gamma-ray emission from rapidly rotating pulsars.
 
 
\end{abstract}

\normalsize

%\newpage

%\nn
 \section{ Introduction}

The \psr system contains a rapidly rotating radio pulsar with spin period $P =
47.76$~ms and spindown luminosity $L_p \simeq 9 \times 10^{35} \ergs $ orbiting
around a massive Be star companion in a highly elliptical orbit with orbital
period $\sim 3.4$~yrs
at an estimated distance $d \simeq 2$~kpc
 %(refs. \cite{johnston1,johnston2,johnston95}).
(Johnston et al. 1992, 1994, 1995; Manchester et al., 1995). 
Table~1 summarizes the main pulsar and orbital characteristics
of the \psr system
%The \psr system 
which is the only known binary where  the interaction of a
relativistic pulsar wind
%$^{\cite{Kennel,hoshino}}$ and
and a mass outflow from a Be star
%$^{\cite{waters2}}$ 
can be studied in detail.
For its characteristics, the \psr system turns out to be
% unique,
%and 
an important astrophysical laboratory for the study of pulsars
interacting with gaseous environments.
%and of their emissions.
Be stars  are characterized by large mass outflows mainly concentrated in
the equatorial plane of the massive star
(e.g., Waters \etal 1988)
%$^{\cite{waters2}}$
 and \psr is expected to
 interact with the Be star outflow especially near periastron.
The nature of the pulsar/outflow interaction can be studied
with the help of  the  high-energy emission (UV, X-rays, \ggg-rays) expected by
the disruption of the Be star outflow and possible formation of
shocks (Tavani, Arons \& Kaspi, 1994; hereafter TAK94).
%$^{\cite{tavaniak}}$.
For the first time, the \psr system allows to study a 
  strong pulsar/outflow interaction. 
 
% There  are several  reasons to justify an extended high-energy  observation
% of the \psr system near periastron. We consider here the main
% possible mechanisms of h
High-energy emission from the \psr system
near periastron can  be produced by several different mechanisms:

{\bf (1)}  {\it Accretion} onto the surface
of the neutron star can occur with
consequent bright X-ray/hard X-ray  emission
for large mass outflow rates  and
high density of material near periastron.
An example of direct accretion is 
the 69~ms X-ray pulsar A0538-66 in the LMC which occasionally
 accretes with an X-ray luminosity   near the Eddington limit
$L_E \sim 10^{38} \ergs$ 
%(ref. \cite{skinner})
(Skinner \etal 1982).
 % demonstrates the ability of mass outflows
% from massive star companions to cause accretion.

{\bf (2)} {\it Propeller effect} emission, caused by gaseous
material being temporarily trapped between the light cylinder and
corotation radius of the pulsar (e.g., Illarionov \& Sunyaev, 1975).
%Campana \etal, 1994);

{\bf (3)}  {\it Pulsar/outflow interaction} caused by the shock
of a relativistic pulsar wind in the mass outflow from the
companion star can produce unpulsed high-energy emission 
(TAK94; Tavani \& Arons, 1997, hereafter TA97). 
The pulsar radiation pressure (caused by a 
MHD
%magnetohydrodynamical
wind of relativistic particles,  electrons, positrons and
possibly ions outflowing from the pulsar)
can be large enough to withstand the compressing ram pressure
of the Be star gaseous outflow
 near periastron.
High-energy shock emission in
pulsar binaries  
is currently 
 poorly constrained, and before the discovery of
\psrp,  only the Crab nebula provided an astrophysical system
unambigously powered by a pulsar wind  %$^{\cite{Kennel}}$. 
(e.g., Kennel \& Coroniti, 1984).
The  \psr  system is interesting in  that it provides
an environment where the pulsar-driven shock can be studied in
hydrodynamical and radiative regimes totally different from the Crab
nebula
%$^{\cite{tavaniak,ta95}}$. % 
(TAK94).

{\bf (4)} {\it Pulsed magnetospheric emission}.
Several rapidly spinning radio pulsars have been detected
at \ggg-ray energies (e.g., Thompson \etal 1994, 
hereafter Th94), and the \psr
parameters   are not dissimilar to those of other \ggg-ray pulsars.
The  pulsar age of \psr is relatively large ($\tau_p \equiv P/2\, \dot{P}
\sim 3.3\cdot 10^5$~yrs) and the inferred dipolar 
magnetic field is sufficiently
small ($B \sim 3.3\cdot 10^{11}$~G) to allow  an interesting comparison
of the detection (or lack thereof) of pulsed high-energy emission from
\psr compared to other  pulsars of similar age (Th94).

We have obtained data from all four CGRO instruments during a 
3-week uninterrupted observation of \psr (January 3-23, 1994)
which included the periastron passage of January 9, 1994.
%eg (We note that this observation is among the longest ever
% carried out by CGRO   for any source).
The faintness of the expected high-energy emission near periastron
justified a long observation.
The length of the observation was crucial in detecting an unambiguous
signal from the source.
The CGRO observation was carried out  during an extended `eclipse'
of the pulsed radio signal from \psr (Johnston \etal 1996;
hereafter J96).
The absence of detectable pulsed  radio emission near periastron
clearly indicates the presence of circumbinary dispersing/absorbing
material in agreement with the expected radial structure of the
Be star outflow.
%Fig.~1 gives the schematic range of orbital phases covered
%by the CGRO observation.
%Table~2 gives the characteristics of the near periastron orbit.

\vskip .1in
%\begin{table}[h]
\begin{center}
{\sl Table 1$^*$}
%\caption{Astrometric, spin, and orbital parameters for \psr$\!\!$.}
\label{ta:pars}
\begin{tabular}{lc}
 \hline\hline
\multicolumn{2}{c}{Astrometric, Spin, and Radio Parameters$^{\ast}$}\\\hline
Right Ascension, $\alpha$ (J2000) & 13h 02m 47.68(2)s \\
Declination, $\delta$ (J2000) & $-$63\deg 50' 08".6(1) \\
Dispersion Measure, DM & 146.75(8) pc~cm$^{-3}$\\
Period, $P$  & 47.762053919(4) ms\\
Spin-down Rate, $\dot{P}$ & 2.2793(4) $\times 10^{-15}$ \\
Period Epoch & MJD 48053.44 \\
% Spin-down Age, $\tau$  & 3 $\times 10^5$ yr\\
% Magnetic Field, $B$  & 3 $\times 10^{11}$ G\\
Spin-down Luminosity, $L_p$  & $8 \times 10^{35}$ erg s$^{-1}$ \\ \hline
\multicolumn{2}{c}{Orbital Parameters}\\\hline
Orbital Period, $P_{\rm b}$ & 1236.79(1) dy \\
Projected semi-major axis, $a_{\rm p}\sin i$  & 1295.98(1) lt~s\\
Longitude of periastron, $\omega$ & 138\deg.6548(2) \\
Eccentricity, $e$ & 0.869836(2) \\
Periastron Epoch, $T_0$ & MJD 48124.3581(2) \\ \hline
\end{tabular}
\end{center}
\label{table:psr}
$^{\ast}$ From Johnston \etal (1994).
 % \scite{jml+94b}.
%\end{table}

%\vskip .3in
%\nn
\section{ Detection of unpulsed emission}

%\vskip .1in
%\nn
\subsection{ OSSE analysis}

OSSE is the only CGRO instrument which detected a positive flux from
the field surrounding \psr (Grove \etal 1995).
%NaI(Tl)--CsI(Na) phoswich detector systems (Johnson et al. 1993).
%It covers the energy range from 50 keV to 10 MeV
%with good spectral resolution.  A
The OSSE  tungsten slat
collimator defines the $3.8$\deg $\times$ 11.4\deg field of view,
which was chosen as a compromise allowing high sensitivity to both
diffuse emission and point sources.  Spectra are accumulated in a sequence of
two-minute measurements of the source field alternated with two-minute,
offset-pointed measurements of background.  
The pulsar is located in the galactic plane at 304.2$^{\circ}$ 
longitude, where
galactic diffuse continuum emission is 
detectable by OSSE.  To minimize the effect of the galactic
emission, the observation of \psr was performed with the
long axis of the collimator oriented perpendicular to the plane and centered
on the pulsar.  Background fields on either side, centered at
308.7$^{\circ}$ and 296.4$^{\circ}$ longitude, were selected to
avoid nearby point sources such as the X-ray binaries GX301--2, Cen X--3,
and 2S~1417-624.  
%jeg5 added 2S1417
%jeg-apj1   restored confusing-source discussion here
These objects, shown by the BATSE instrument on CGRO to be occasional
emitters in the 20--50 keV band,
% (R. Wilson, private communication),
are each suppressed by a factor of $>10^2$ by the collimator and
account for no detectable emission in the source field.  Unavoidably, 
the source field contained the Be-star binary pulsar GX304--1; however, 
this X-ray pulsar has not been detected by BATSE in three years of daily
observation.
% (M. Finger, private communication), 
%dt and it was shown by
%EXOSAT to be in an off state in 1984
%$^{\ci{Pietsch}}$.
 % (Pietsch et al. 1986).
A total exposure time $T = 1.9 \times 10^5$ s of highest-quality data
were collected in two OSSE detectors on the \psr source field, with
approximately an equal time on the background fields.  
%mt33 suppressed sentence here, too detailed for Nature
%Two of the
%detectors showed an exceptionally high rate of small, short-term gain
%variations and have been eliminated from the following analysis.
%Including these two detectors does not substantially alter the detection
%significance (see below), but the gain variations can confuse the
%spectral interpretation.

%\section {Results}

The 
%eg energy spectrum 
photon number spectrum 
from the \psr system  is shown in Fig.~1. % \ref{spec}.
Emission at the level of $\sim 4$~mCrab is detected by OSSE 
 from 50 keV to $\sim$200 keV, with a total statistical
significance of 4.8$\sigma$.
%mt5 p. 5, first paragraph of Results, second sentence:  Add after 4.8 sigma, "
%at a level of $\sim$5~mCrab flux units.
%
%jeg34 in the two 
%mt33 added 'most' 
%jeg34 most stable detectors
%(4.3$\sigma$ including all four
%detectors).  
%mt33 rearranged paragraphs and sentences here is a more coherent pattern
The best-fit power law photon number spectrum for photon energies
$E$ in the  50--300 keV band  is
$d\, N_{\gamma} / d \, E = (2.8 \pm 0.7) \times 10^{-3} (E/{\rm 100 \,
keV})^{-\alpha}$
ph.~cm$^{-2}$ s$^{-1}$ MeV$^{-1}$,
with photon index $\alpha = 1.8 \pm 0.6$.
%and $E_{2} = E/(100$ keV).
The power-law fit is good ($\chi_{\rm dof}^2 = 0.90$ for 40 degrees of freedom).
Also shown in Fig.~1
% \ref{spec} 
is the ASCA spectrum for the January 26, 1994 observation 
(Kaspi et~al. 1995; hereafter KTN95).
%jeg-apj2 added uncertainty in extrapolation
%We note that the uncertainties in ASCA's fit parameters are such that the three
%models are essentially equivalent at 100 keV.
The emission in the OSSE band is  consistent in intensity and spectral
%jeg-apj2 'single' > 'simple', removed 'higher state'
shape with a simple power-law extrapolation of the ASCA
% higher state of the ASCA 
observation.  The continuation of the power law spectrum to
hard X-ray energies is very significant for the interpretation of
%jeg-apj2 added "see Discussion"
the emission.
% (see Discussion).
  At 2.0 kpc (Johnston et al. 1994), 
the inferred 50--200 keV
luminosity of the \psr system is $L_X \simeq 3 \times 10^{34}$ erg s$^{-1}$, 
and the hardness of the OSSE spectrum suggests that the luminosity per 
energy decade peaks near or above 200 keV.
%

%\par
There is no evidence for variability in the OSSE data; however, because of
the relatively low statistical significance of
the detection, the constraints on variability are not stringent.
The 95\%-confidence upper limits on 50--200 keV variability on daily
or weekly timescales are $\sim$190\% and $\sim$70\%, respectively.
Any flux decrease by $\sim$50\% during the 3-week observing period,
%jeg-apj1 added: as suggested by ASCA
as suggested by the ASCA data, is undetectable by OSSE.
% {\bo 
% %p. 5, second paragraph of Results:  Add sentence at end, "
% Because of the
% brevity of our observation in relation to the 3.4-year binary period, we are
% unable to place a meaningful limit on modulation at orbital timescales. }
%\vskip .05in

For weak detections such as that from the \psr field,
it is important to address
possible systematic errors or alternate sources for the emission.
A study of residuals in the 50--200 keV band from an aggregrate of many
OSSE pointings indicates that the systematic errors are smaller than the
statistical errors and, therefore, that
the $\sim$5$\sigma$ excess reported here should be exceedingly rare.  
The background fields of view were chosen to exclude known
point sources of hard X-rays in the region.
%, 
% and the source field otherwise contained only GX304--1, which is
% apparently in an off state (see Observations).
% %jeg-apj4 and 
%undetected by BATSE in three years of daily observation (M. Finger,
%private communication).
% Furthermore, we find no evidence for pulsations from GX304--1 in
% the OSSE data at periods between 267 s and 275 s, which reasonably
% bound extrapolations of the 
%historical period.	%jeg-apj4 'extrapolation' added
%
%jeg-apj4  slight mods to following paragraph
%
% OSSE is conducting an extensive survey of the galactic diffuse line and
% continuum emission (e.g. Purcell et al. 1994).  To date, measurements
% appropriate to estimating the
% diffuse emission in the \psr region, i.e. those at similar separation from
% the galactic center and without significant contributions from known point
% sources, are available only on the opposite side
% of the galactic center, at longitudes 
% 40$^{\circ}$ and 58$^{\circ}$. 
 From  measurements near the galactic plane, and for  a simple one-dimensional
galactic plane model, we
estimate that the large-scale (i.e. $\sim10^{\circ}$) longitudinal gradient
in the emission 
per degree in the 50--200 keV band is
$\sim -3 \times 10^{-5}$ ph.~cm$^{-2}$ s$^{-1}$ MeV$^{-1}$ deg$^{-2}$.
Assuming that the diffuse emission is 
%generally 
symmetric about the
galactic center and convolving this emission through
the instrument response, we estimate that the residual diffuse 50--200 keV
flux for our viewing strategy of the \psr region is
$\sim 2 \times 10^{-4}$ ph.~cm$^{-2}$ s$^{-1}$ MeV$^{-1}$,
a factor of $\sim$10 less than the observed flux.
We note, however, that the
level of small-scale, local fluctuations in the diffuse emission is
somewhat uncertain and not easily measured, 
given the presence of nearby point sources.  Such
fluctuations are the most
likely alternative explanation for the observed flux.

%\vskip .2in
%\nn
\subsection{ BATSE analysis}

BATSE data were analyzed  near the time of periastron using  the standard
BATSE occultation technique.
The resulting  light curve spans 
the period of truncated Julian days  (TJD) 9340-9380.
No significant flux was detected, with an upper limit of
$\Phi_{BATSE} \loe  0.02 \, \rm ph.~cm^{-2} \, s^{-1}$
in the 20-120~keV  energy band. This limit was obtained assuming
a power law index of emission $-2$ in the deconvolution.
This upper limit is  ultimately a consequence of systematic
effects which are observed as a scatter of the flux around a
zero mean value.
We notice that $\Phi_{BATSE}$ is about one order of magnitude larger
than the detected flux by OSSE.

%\vskip .2in
%\nn
\subsection{COMPTEL analysis}

COMPTEL data for  \psr from the VPs 314-316
were analyzed in four different energy channels 
% 0.75-1,1-3,3-10,
at 1, 1-3, 3-10,
10-30~MeV. 
A spatial analysis was performed with the use of the 
maximum likelihood method to the three dimensional COMPTEL
data space, which is defined by the coordinates of the scatter 
direction of the incoming photon in the upper-detector layer
and by the scatter angle (Schoenfelder \etal 1993).
The dominating background distribution  in this data space
has been determined  applying a filter technique to the 
measured count distribution in the same data space.
This approach produces a smooth background distribution in
which source-like enhancements are suppressed.
In order to verify that the background treatment does not
suppress evidence for a genuine source signal at the position
of \psrp, we applied three different approaches to the 
background generation.
%No evidence for a detection of the \psr system was found.
% Table~2 gives the $2-\sigma$ upper limits to the source
% fluxes for the four chosen energy channels.
No evidence for a detection of the \psr system
was found in the COMPTEL data.

%\vskip .2in
%\nn
\subsection{ EGRET analysis}

We performed a spatial analysis of EGRET data using a
maximum likelihood technique (Mattox \etal 1996).
%wh Neither the spatial nor the timing analyses yielded a
% positive signal for any of the energy ranges or subsets
% of the data.
The observed intensity from the direction of \psr was
compared with that one expected from the galactic and
extragalactic diffuse \ggg-ray emission as modelled
by Hunter \etal (1996).
Upper limits (95\% confidence)
were derived from the  likelihood  analysis.
We have also examined the EGRET data for \ggg-ray transient
emission from the \psr system on time scales ranging from
hours to several days. No significant transient emission was detected
for photon energies $E_{\gamma} > 80$~MeV.
%
 %analyses, using techniques previously  applied to EGRET data
%(Mattox \etal 1996, Th94).  % Thompson \etal 1994).  %; de Jager, 1994).
The most stringent constraints on high-energy  radiation
are provided by the spatial analysis for the whole data set 
(see Table~2).
% These limits are given in Table~3, where the conversion from
% photon number to energy is done assuming a differential spectrum
% proportional to $E^{-2}$.

\begin{figure*}[thbp]
 % \picplace{4cm}
 \centering{ \vspace*{-.5cm}
 \psfig{figure=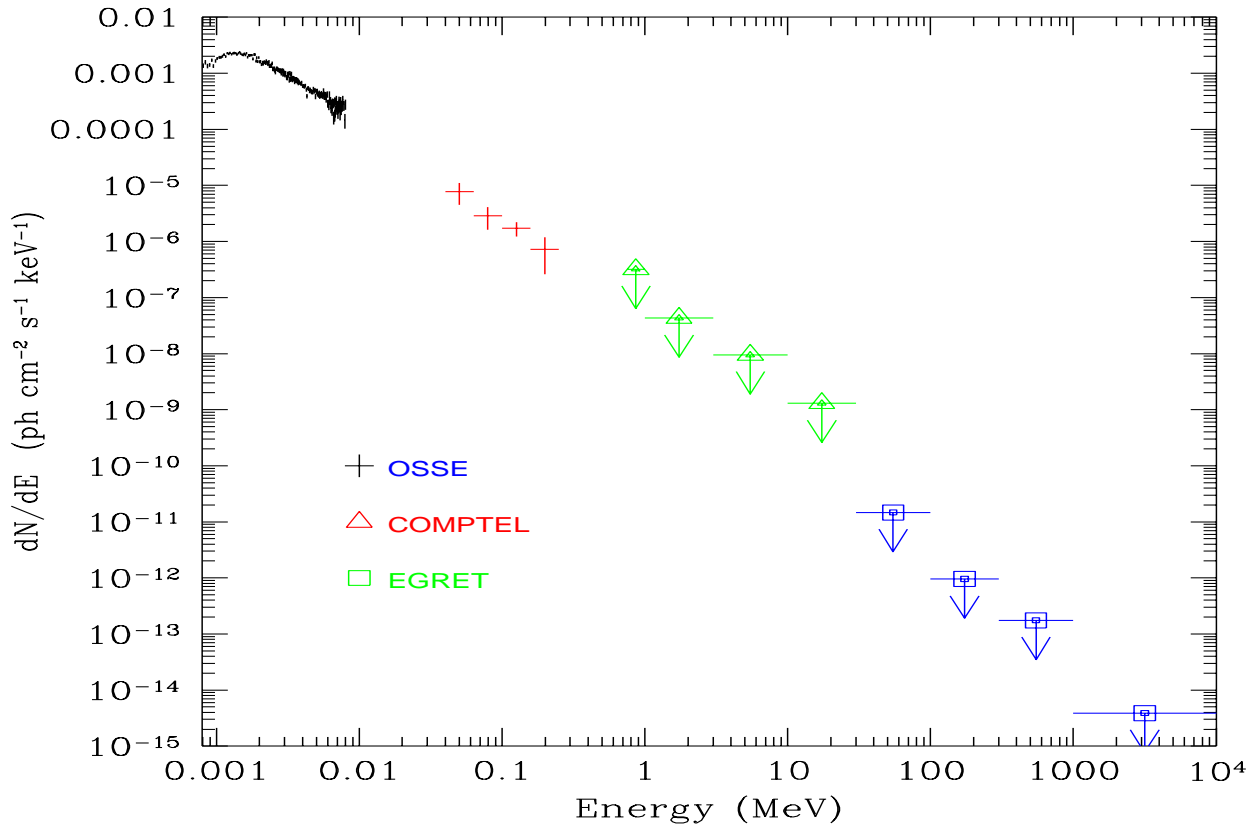,height=12.cm,width=17.cm} }
 \caption{
GRO multi-instrument spectrum of \psr near periastron
during the period January 3-23, 1994 (data from Grove \etal
1995)
% ; Tavani \etal 1995). 
BATSE upper limits are not given, and
they are a factor of $\sim 10$ larger than the OSSE data points.
 Also ASCA data (0.5-10~keV)
 obtained during the post-periastron
January 26, 1994 observation are given for comparison
(KTN95).
%(Kaspi \etal 1995). 
}
 \label{Figure1}
\end{figure*}

%\vskip .3in
%\nn
\section{ Search for pulsations}

Using the radio ephemeris of Table~1, we searched for pulsed emission
at the rotation period of the pulsar calculated for   the
particular time of observation.
The orbital motion introduces non-negligible Doppler shifts of the
photon arrival times, and standard techniques were used to 
correct the observed photon arrival times to the pulsar
barycenter.
No pulsed emission was detected by any CGRO instument.
However, we caution that because of the long integration
required in the \ggg-ray regime, small errors in the pulsar
ephemeris could in principle  smear  coherent pulsations.
Due to the relatively small flux detected from the \psr
system, the best CGRO upper limits on pulsed emission were
obtained in  the OSSE energy range, corresponding to
$\eta_{1259} \loe \eta_{Crab}$, where $\eta$ is the efficiency
of conversion of spindown energy into pulsed  \ggg-ray emission.

%\vskip .2in
%\nn
%\subsection{ OSSE analysis}
%\vskip .2in
%\nn
%\subsection{ BATSE analysis}

The OSSE 95\% confidence upper limits for pulse widths of
0.15 or 0.5 in the 50-180~keV band during the  21-day observation
are $\Phi_{0.15} = 3\cdot 10^{-4}$ and 
$\Phi_{0.5} = 6\cdot 10^{-4} \rm \, ph. \, cm^{-2} \,
s^{-1} \, MeV^{-1}$, respectively.
The former limit corresponds  to $\sim 3\cdot 10^{-3}$ 
Crab pulsar flux units.

For pulse periods less than 1~s, BATSE data can be acquired
in the `folded-on-board' (FOB) mode.
In this mode, data in the appropriately pointed Large Area
Detectors  undergo
pre-telemetry folding at a specified period.
FOB data, folded at the pulsar spin period, was obtained for
the time intervals  TJD~9341-9375 and TJD~9397-9433.
The time coverage extends from approximately 20 days before
periastron to 72 days later.
In none of the folding time intervals  we find a significant deviation
from a flat light curve.
We performed a Monte Carlo calculation to determine an upper limit of
$\sim 1.5 \times  10^{-3} \rm \, ph. \, cm^{-2} \, s^{-1}$
for Gaussian-shaped  pulsed emission
in the energy range 20-100~keV.
% typically of value 
% $\sim 1.5 \times  10^{-3} \rm \, ph. \, cm^{-2} \, s^{-1}$
 % Table~4 gives the 
% derived upper limits for Gaussian-shaped light curve of 
% pulsed emission.

%\vskip .2in
%\nn
%\subsection{ COMPTEL analysis}

We carried out a timing analysis of COMPTEL data for the same energy
channels used for the study of unpulsed emission.
 The applied event selections are based on
experience gained in the analyis of  the Crab and Vela pulsed
signal. The data have been folded using the parameters of two
different  ephemerides (Johnston \etal 1994; J96).
One for pre-periastron data, and one fitting all radio data up to 
May 1994. However, 
it should be realized that there is an uncertainty in the instantaneous
phase during the periastron passage, with  
shifts possibly as large as several milliseconds.
The derived phase histograms, for both ephemerides and energy
intervals, do not show sufficient evidence for a  pulsed signal.
%Fig.~3 shows the distributions applying the post-periastron
%ephemerides in the four energy intervals.
%No significant pulsed emission is detected.

%\vskip .2in
%\nn
%\subsection{ EGRET analysis}

Timing analysis of EGRET data in different energy intervals did not yield a
positive detection of \psrp. 
Applying the technique of Thompson et al. (1994), upper
limits are obtained in a search for a pulsed signal assuming a sinusoidal
shape, as well as assuming a narrow pulse shape. The latter approach produces
somewhat lower upper limits (see Table 2).
% 
% Table~2 gives EGRET upper limits which are not dissimilar from those
% obtained in a search of a  pulsed signal  of sinusoidal
% shape.
% A search performed assuming a narrow pulse shape is expected
% to produce somewhat smaller upper limits than those of Table~2.

%\begin{table}
\begin{center}
%\caption{
{\sl Table 2: EGRET   95\% upper limits}
% \begin{tabular}{|l|c|c|c|c|}
\begin{tabular}{lcccc}
\hline
% Energy Range  & Photon Limit & Energy Limit \\
% MeV   &    ph/cm$^{-2}$s      &   ergs/cm$^{-2}$s     \\
 Energy Range (MeV) & $F^{(a)}$ &  $F_E^{(b)}$ & $F_{np}^{(c)}$ &$F_{bp}^{(d)}$\\
% Photon Limit & Energy Limit & Narrow pulse & Broad pulse\cr
%(MeV)   &  ($10^{-8} \, \rm cm^{-2} s^{-1}$)  &
%($10^{-11}\,  erg cm^{-2} s^{-1}$)  & 
%ph/cm$^{2}$s & ph/cm$^{2}$s\cr
%($10^{-8} \, \rm cm^{-2} s^{-1}$)  &($10^{-8} \, \rm cm^{-2} s^{-1}$)\\ 
\hline
30-100& 79.  & 6.6  & 100. & 250. \cr
100-300& 18. & 4.8 & 20.  & 49.  \cr
300-1000& 9.4 & 7.8 & 9.6 & 23.  \cr
1000-10000& 4.5 & 18.  & 7.3 & 21. \cr
%30-100& 79.  & 6.6  & 100.   & 250. \cr
%100-300& 18. x 10$^{-8}$ & 4.8 x 10$^{-11}$  & 20. x 10$^{-8}$  & 49. x 10$^{-8}$ \cr
%300-1000& 9.4 x 10$^{-8}$ & 7.8 x 10$^{-11}$  & 9.6 x 10$^{-8}$ & 23. x 10$^{-8}$\cr
%1000-10000& 4.5 x 10$^{-8}$ & 18. x 10$^{-11}$  & 7.3 x 10$^{-8}$ & 21. x 10$^{-8}$\cr
%}}$$
%
 %30-100& 79. x 10$^{-8}$ & 6.6 x 10$^{-11}$ \\
 %100-300& 18. x 10$^{-8}$ & 4.8 x 10$^{-11}$ \\
 %300-1000& 9.4 x 10$^{-8}$ & 7.8 x 10$^{-11}$ \\
 %1000-10000& 4.5 x 10$^{-8}$ & 18. x 10$^{-11}$ \\
\hline
\end{tabular}
%\caption{EGRET   95\% upper limits}
 %\label{table:egret}
\end{center}
\vskip -.03in
\noindent
{\small 
$(a)$ Photon flux  upper limit (in units of $10^{-8} \, \rm cm^{-2} s^{-1}$);
$(b)$ energy flux upper limit (in units of $10^{-11}\, \rm erg \, cm^{-2} s^{-1}$);
$(c)$ flux  upper limit of pulsed emission for a narrow pulse
(in units of $10^{-8} \, \rm cm^{-2} s^{-1}$);
$(d)$ flux  upper limit of pulsed emission for a broad  pulse
(in units of $10^{-8} \, \rm cm^{-2} s^{-1}$).
The  narrow and broad pulse light curves have been assumed to be
Gaussian-shaped 10\% wide in phase and  a sinusoid, respectively.}
%\end{table}

\vskip .08in

%wh We notice that b
By adopting an integrated energy upper limit
($E_{\gamma} > 100$~MeV) $\Phi_{EGRET} \sim 10^{-11} \,
\rm erg \, cm^{-2} \, s^{-1}$ and a beaming factor $f_b=1$~sr,
we obtain an upper limit to the efficiency $\eta$ in the EGRET
energy range $\eta'_{EGRET} \loe 0.005$.
This upper limit is about a factor of 70 larger than in the 
case of the Crab in the same energy range.
For comparison, we note that
in the EGRET energy range
the efficiencies for pulsed emission for the Vela  pulsar
and PSR~1706-44 are 
$\eta_{Vela} \sim 0.002$, and
$\eta_{1706-44} \sim 0.007$, respectively.
We conclude that the current EGRET upper limits on pulsed
emission from \psr exclude a relatively large efficiency as for
 PSR~1706-44.
Fig.~2  shows  the pulsar parameter space with the inclusion
of \psr and with the EGRET \ggg-ray pulsars clearly marked.
We notice that \psr has parameters typically smaller by a factor
of $\sim 2$ than those of PSR~1951+32 recently revealed by
EGRET with an efficiency $\eta_{1951+32} \sim 0.0038$
(Ramanamurthy \etal 1995).

%\vskip .3in
%\nn
\section{Comparison with radio and ASCA data near periastron}

Extensive radio observations of \psr carried out in 1993 and
1994 (J96)
showed a first sign of pulsar/outflow interaction in the 
\psr system in October 1993 (approximately at
${\cal T}-94$,
where ${\cal T}=$~January 9.2, 1994, TJD~49361.7  is the 
periastron date).
Substantial   radio pulse depolarization was observed,
with occasional eclipsing behavior at 1.5 GHz at the end of
November 1993. The pulsar was not detected in 1.5 GHz data on
Dec. 20, 1993 (${\cal T} -20$) and reappeared  on
Feb. 4, 1994 (${\cal T} + 24$). During the 44-day eclipse
\psr was not detected in  extensive
observations at 1.5 and 8.4 GHz (J96).
The dispersion measure showed substantial increase in the pre-eclipse
data, reaching a value of $\Delta \, DM \sim 11 \, \rm cm^{-3}
\, pc$ above normal,
corresponding to an average free electron density of 
$N_e \sim 3.3\cdot 10^{19} \rm \, cm^{-2}$.
By mid-April 1994, both the linear polarization and the rotation
measure were  back to pre-eclipse values.
Extensive ASCA observations of the \psr system near and after periastron
(see Fig.~1; KTN95, Hirayama \etal 1996, TA97)
%Tavani \etal 1995) 
show a non-thermal
X-ray spectrum with luminosity $L_X \sim 10^{34} \rm
\, erg \, s^{-1}$ and photon index $1.9-1.6$.
The deduced column density to photoabsorbing  atoms
is particularly low $N_H \simeq 5\cdot 10^{21} \rm \,
cm^{-2}$, and consistent with neglibile intrinsic absorption.
Furthermore, $N_H$ was observed to be constant throughout the
whole set of ASCA observations covering the periastron passage
of \psrp.
A crucial ASCA observation of \psr  carried out on
Feb. 28, 1994 (when pulsed radio emission from \psr was again visible, J96)
gave X-ray intensity, spectrum and column density similar
to those previously observed near periastron (Hirayama et~al. 1996).
We conclude that both the radio and the X-ray data on \psr
are consistent with a relatively `dilute' nebular environment
surrounding \psr near periastron and in sharp contrast with enviroments
surrounding accreting systems.

%\vskip .3in
%\nn
\section{ Theoretical discussion}

Hard X-ray emission with a photon index $\alpha \sim 1.8-2$ is a
manifestation of energetic particle acceleration.
% Accreting neutron stars do not usually emit in the hard X-ray range
% (e.g., Barret \& Vedrenne, 1994)
% and only sporadic hard X-ray emission from a few X-ray bursters
% has been reported with 
% a power-law fit to the spectrum typically of 
% photon index $\sim 2-3$ or larger.
Accreting neutron stars tend to be episodic emitters in the 
hard X-ray band typically
 with soft thermal or steep power law (photon index ~3) 
spectra (e.g. Barret \& Vedrenne 1994).
In the case of emission from the \psr system, 
one can argue against accretion or propeller mechanisms  because of:
(A)  lack of X-ray/\ggg-ray pulsations, 
(B)  absence of significant X-ray fluctuations  
within timescales  of order of days or less (KTN95),
(C)  relatively low X-ray luminosity 
and negligible absorption column density
(TA97),
(D) hardness of the power-law X-ray spectrum up to $\sim 200$~keV.
%(D) spectral hardness which is  substantially
%different  than for accreting neutron stars 
%(and consistent with an extrapolation
%to soft X-ray energies).
% all argue against accretion-powered emission 
%(Tavani \& Arons 1995, KTN95).
%jeg-apj4 { (TA95, KTN95)}.
For the parameters of the \psr system, 
it can be shown that accretion or propeller
regimes near periastron  are unlikely to occur 
except for unusually large
values of the  the mass loss rate from the
Be star companion  (TA97).

Hard X-rays provide crucial information on the
shock emission due to the interaction of 
the relativistic particle wind
of \psr with the nebular gas of the Be star mass outflow. 
As in the case of the Crab nebula 
(Kennel \& Coroniti 1984, Gallant \& Arons
1994),
shock acceleration and synchrotron 
radiation capable of producing unpulsed
hard X-ray emission occurs at a shock radius within the \psr binary
where pressure balance is established (TAK94)
%avani et al. 1994) %jeg-apj4 TAK94)
between the pulsar
relativistic wind (electrons/positrons, possibly ions) and the 
ram pressure of the Be star outflow.
The radiative environment near the  shock radius  
within the \psr system is
quite different from that 
of the Crab nebula, and the large synchrotron and inverse 
%jeg5 moved eqn
Compton cooling rates ($\sim 10^{-2}-10^{-3} \, \rm s^{-1}$)
at the \psr periastron make the detection of high-energy 
emission a unique diagnostic of
shock acceleration subject to strong radiative cooling. % (TA95).
%Tavani et al. 1994,
%Tavani \& Arons 1995).  %jeg-apj4 {\mt (TAK94, TA95)}. 
By comparing the CGRO and ASCA results 
%mt5(Kaspi et al. 1995), %jeg-apj4 (KTN95),
a unified picture for the emission and spectrum emerges.
%mt5
{\bo
Spectral calculations for models with a weak ram pressure of the Be star 
outflow show that the dominating inverse Compton spectrum is typically
much too flat to agree with the 
OSSE data (TA97). } %avani \& Arons, 1995). }
%jeg5 slight changes in word order
{\bo

The spectral shape and intensity of the power-law emission
from the \psr system are consistent with synchrotron radiation
of shock-accelerated electron-positron pairs
for an intermediate  value of the ram pressure of the Be star
outflow and a shock distance relatively close to the pulsar
rather than to the Be star.
The inferred characteristics of the electron/positron pairs in the
pulsar wind are
a Lorentz factor of $\gamma_1 \sim 10^6$ 
and a ratio of electromagnetic
energy density to particle kinetic energy density in the 
wind of $\sigma \sim 10^{-1}-10^{-2}$ (TA97). }
%(Tavani \& Arons, 1995). }
The observed  efficiency of conversion of the \psr
pulsar wind kinetic and electromagnetic
energy into radiation in the 50-200~keV band is 
$\sim 3 \%$, i.e., a value
comparable to the efficiency of the Crab nebula.
%for a system distance of 2~kpc (Johnston et al. 1994).
The absence of a spectral break from the soft X-ray band above
the photon energy 
$\epsilon_{min} \sim 1$~keV 
%(Kaspi et al. 1995) 
through  %jeg-apj4 (KTN95) through
the lower limit on the high-energy cutoff, $\epsilon_{max}
\goe 200$~keV from the OSSE detection,  and the  COMPTEL and EGRET
upper limits strongly constrains the
acceleration mechanism.
The collisionless pulsar wind shock must produce a power-law
particle distribution function of index $\delta$
between a minimum \eee-pair energy $E_1 = \gamma_1 \, m_e \, c^2$ and  
an upper energy cutoff $E_c = \gamma_{c} \, m_e \, c^2$, 
with $m_e$ the 
electron-positron mass and $c$ the speed of light.
For a synchrotron model of shock emission, we 
obtain the interesting constraint (TA97)
\be \frac{\gamma_{c}}{\gamma_1} \goe
\left( \frac{\epsilon_{max}}{\epsilon_{min}} \right)^{1/2} \sim 10  \;\;,  \en
where $\epsilon_{max} \simeq 200$~keV.
Furthermore, we obtain  $\delta = 2 \, \alpha -1 \sim 2-3$.
A spectral cutoff is likely to
occur near 1-10~MeV, in agreement with the calculated
 spectra for pulsar wind termination shocks subject to 
strong radiative cooling
(TA97).
The CGRO observation of the \psr system therefore demonstrates, 
for the first
time in a galactic source,
the ability of  shock  acceleration to efficiently
energize a relativistic plasma subject to radiative cooling of typical
timescales  $\tau_c \sim 10^2-10^3$~s.
The important derived constraints on the radiation 
efficiency ($\sim 3\%$), 
acceleration timescale ($\loe \tau_c$), index of the post-shock
particle distribution ($\delta \sim 2-3$) are of 
crucial interest for
any theory of relativistic shock acceleration.

\begin{figure}[thbp]
 % \picplace{4cm}
 \centering{ 
%\vspace*{-.5cm}
 \psfig{figure=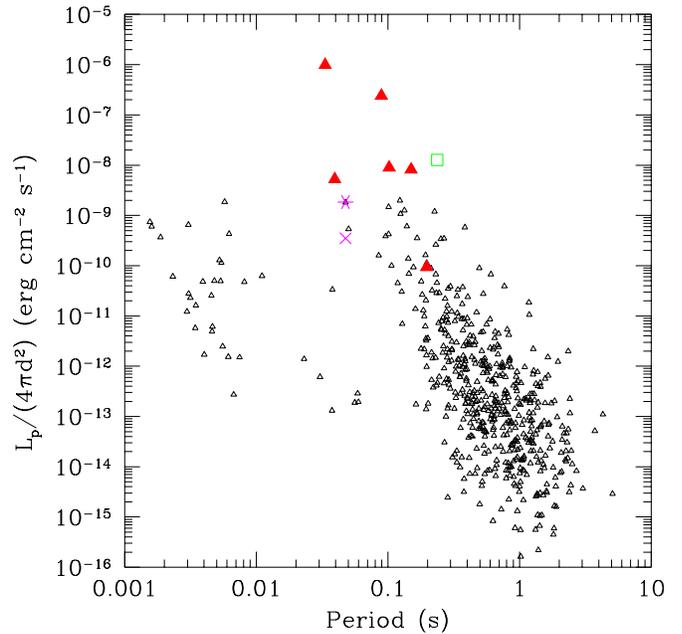,height=9.cm,width=9.cm} }
 \caption{
Distribution of pulsar spindown energy flux vs. pulsar spin period for
706 pulsars. 
 $L_p$ and $d$ are the pulsar
spindown luminosities and distances (Taylor \etal 1993).
 Large triangles indicate detected
 gamma-ray pulsars (Thompson \etal 1994). 
Geminga parameters (for an assumed distance of 150~pc)
 are marked by an open square.
The star  and the tilted cross
mark  the parameters of \psr for $d=2$~kpc (J96)
 and $d=4.6$~kpc (Taylor \etal, 1993), respectively.
}
 \label{Figure2}
\end{figure}

% \vskip .3in
% \nn
% {\LARGE \bf Conclusions}

% Our results can have a broad range of applications to astrophysical sources 
% with relativistic plasmas being shocked and accelerated in a strongly radiative
% environment.  Our observations of the \psr system show that
% shock emission is an efficient mechanism of unpulsed high-energy
% radiation in pulsar binaries. Poorly understood  unidentified
% {\bo and time-variable} high-energy galactic sources (e.g., Fichtel et al. 1994)
% may reveal energetic pulsars powering unpulsed emission in binaries. 

\acknowledgements{
We thank S. Johnston and R.N. Manchester for  useful exchange of information,
and F. Nagase and V. Kaspi for discussions and
collaborative effort on the ASCA data. Research supported by NASA grant
NAG5-2234.}
% (MT) and NAGW 2413 (JA), NSF grant AST 91-15093 (JA) and NASA
%contract S-10987-C (OSSE). }

\normalsize

%\vskip .3in
%\nn
\section{ References}

\renewcommand{\apj}{ApJ}
\newcommand{\mnras}{MNRAS}

\rref{Barret, D. \& Vedrenne, G. 1994, ApJS, 92, 505 }
%\rref{Fichtel, C.,E.,  et al. 1994, ApJS, 94, 551}
\rref{Gallant, Y.A., \& Arons, J. 1994, \apj, 435, 230}
\rref{Grove, J.E., et al., 1995, ApJ, 447, L112}
\rref{Hirayama, M., \etal, 1996, to be submitted to PASP}
% \rref{Hoshino, M., \etal, 1992, \apj, 390, 454}
%Arons, J., Gallant, Y.A. \& Langdon, A.B. 1992,
%   \apj, 390, 454}
\rref{Hunter, S. \etal, 1996, ApJ, in press}
\rref{Illarionov A.F. \& Sunyaev, R.A., 1975, A\&A, 39, 185}
\rref{Johnson, W.N., et al. 1993, ApJS, 86, 693}
\rref{Johnston, S., et al., 1992
% Manchester, R.N., Lyne, A.G., Bailes, M., Kaspi, 
   % V.M., Qiao, G., \& D'Amico, N. 1992 
\apj, 387, L37}
\rref{ Johnston, S., \etal, 1994,
% Manchester, R.N., Lyne, A.G., Nicastro, L. \&
   %Spyromilio, J. 1994, 
\mnras, 268, 430}
\rref{Johnston, S., \etal, 1996, \mnras, 279, 1026 (J96)}
\rref{Kaspi, V., % \etal, 1995, ApJ, in press}
 % Tavani, M., Nagase, F.  \etal, 1995, ApJ, in press (KTN95)}
 Tavani, M., Nagase, F., Hoshino, M., Aoki, T.,
    Hirayama, M.,  Kawai, N. \& Arons, J., 1995,  ApJ, 453, 424 (KTN95)}
\rref{ Kennel, C.F. \& Coroniti, F.V. 1984, \apj, 283, 694}
\rref{Manchester, R.N. \etal, 1995, ApJ, 445, L137}
\rref{Mattox, J., \etal, 1996, ApJ, in press}
% \rref{ Pietsch, W., et al., 1986, A\&ASS, 163, 93}
\rref{ Purcell, W.R., et al., 1994, in AIP Conf. Proc. 304, p. 403}
\rref{Skinner, G.K., \etal, 1982, Nature, 297, 568}
%\rref{ Ray, P.S., et al. 1992, AIP Conf. Proc. no. 280, p. 249}
    % ed. M. Friedlander, N. Gehrels, and D.J. Macomb (New York), p. 249}
\rref{Ramanamurthy \etal, 1995, ApJ, 447, L109}
\rref{Schoenfelder V. \etal, 1993, ApJS,  86, 657}
\rref{Tavani, M., Arons, J., Kaspi, V. 1994, ApJ, 433, L37}  %  (TAK94)}
%\reference Tavani, M., et al. 1995, submitted to \apj
\rref{Tavani, M. \& Arons, J., 1997, ApJ, in press;\\
astro-ph/9610212 (TA97)}
% \rref{Taylor, J.H., Manchester, R.N. \& Lyne, A.G., 1993, ApJS, 88, 529}
\rref{Taylor, J., Manchester, R., Lyne, A., 1993, ApJS, 88, 529}
\rref{Thompson, D., et al., 1994, ApJ,  436, 229 (Th94)} 
\rref{Waters, L.B.F.M., et al. 1988, A\&A, 198, 200}
\rref{White, N.E., Nagase, F. \& Parmar, A.N. 1995, in
   {\it X-Ray Binaries}
%eds. W.H.G. Lewin, J. van Paradijs \& E.P.J.
   %van den Heuvel 
(Cambridge Univ. Press), p. 1}   %, in press}
%\end{references}

\end{document}